\documentclass[journal=cmatex,layout=twocolumn,10pt]{achemso}

\usepackage{natbib}
\usepackage[utf8]{inputenc}
\usepackage{graphicx}
\usepackage{dcolumn}
\usepackage{bm}
\usepackage[breaklinks,pdffitwindow=false,bookmarks=true,pdfauthor={Hoffmann et al.},colorlinks=false,citecolor=false, breaklinks=true,linkcolor=black]{hyperref}
\usepackage{titlesec}
\titleformat{\subsubsection}[runin]{\bfseries\itshape\normalsize}{\thesubsection}{1em}{}
\titlespacing*{\subsubsection}{0pt}{.8em}{1em}

\title{\texorpdfstring{
Multi-Lattice Kinetic Monte Carlo Simulations from First-Principles:\\
Reduction of the Pd(100) Surface Oxide by CO}{
Multi-Lattice Kinetic Monte Carlo Simulations from First-Principles:\\
Reduction of the Pd(100) Surface Oxide by CO}}

\author{Max J. Hoffmann}
\email{max.hoffmann@ch.tum.de}
\affiliation{Chair for Theoretical Chemistry and Catalysis Research Center, Technische Universit\"at M\"unchen, Lichtenbergstr. 4, 85747 Garching (Germany)}

\author{Matthias Scheffler}
\affiliation{Fritz-Haber-Institut der Max-Planck-Gesellschaft, Faradayweg 4-6, 14195 Berlin (Germany)}

\author{Karsten Reuter}
\email{karsten.reuter@ch.tum.de}
\affiliation{Chair for Theoretical Chemistry and Catalysis Research Center, Technische Universit\"at M\"unchen, Lichtenbergstr. 4, 85747 Garching (Germany)}

\begin{document}

\begin{tocentry}
    \includegraphics[width=9cm]{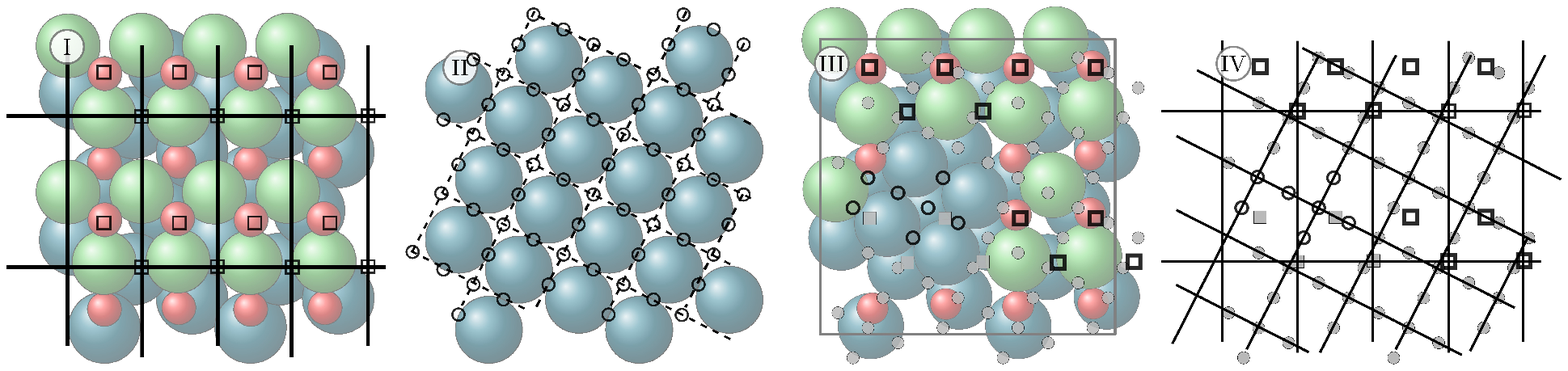}
\end{tocentry}

\begin{abstract}
We present a multi-lattice kinetic Monte Carlo (kMC) approach that efficiently describes the atomistic dynamics of morphological transitions between commensurate structures at crystal surfaces. As an example we study the reduction of a $(\sqrt{5} \times \sqrt{5})R27^{\circ}$ PdO(101) overlayer on Pd(100) in a CO atmosphere. Extensive density-functional theory calculations are used to establish an atomistic pathway for the oxide reduction process. First-principles multi-lattice kMC simulations on the basis of this pathway fully reproduce the experimental temperature dependence of the reduction rate [Fernandes {
\em et al.}, Surf. Sci. {\bf 2014}, {\em 621}, 31-39] and highlight the crucial role of elementary processes special to the boundary between oxide and metal domains.
\end{abstract}

\section{Introduction}

In recent years first-principles (1p) kinetic Monte Carlo (kMC) simulations have established a new standard for the microkinetic modeling of surface chemical processes and heterogeneous catalysis \cite{ratsch_1997,ruggerone_1997,fichthorn_2000,kratzer_2002,reuter_steady_2004, sabbe_first-principles_2012}. On the one hand, this approach employs material-specific rate constants for the elementary processes, presently typically computed by density-functional theory (DFT) and transition-state theory (TST). On the other hand, and in contrast to still prevalent mean-field rate equation based microkinetic models, the 1p-kMC approach explicitly accounts for spatial heterogeneities, correlations, and fluctuations of the atomic structure. The downside of the approach is the computational cost of the underlying DFT-TST calculations. In practice, 1p-kMC simulations are therefore hitherto performed on lattice models. This exploits the translational symmetry of the crystalline lattice to dramatically reduce the required number of different 1p rate constants \cite{reuter_first-principles_2012,hoffmann_kmos:_2014}. For simple reaction networks like the CO oxidation reaction at low-index single-crystal model catalyst surfaces, this number of 1p rate constants can be as low as 10-20, comprising adsorption, desorption, diffusion and reaction steps at and between the different lattice sites \cite{reuter_steady_2004,rogal_first-principles_2007,rieger_effect_2008}.

By construction, the conventional lattice 1p-kMC approach cannot directly be applied to (surface) morphological rearrangements, {\em i.e.} any type of phase transition that includes changes in the considered lattice site arrangement. Notwithstanding, for the (late) transition-metal catalysts typically employed in oxidation catalysis, a possible formation of (surface) oxides has been discussed. Possibly, especially the continuous formation and reduction of such oxidic films could be a major actuator for the observed catalytic activity \cite{michaelides_when_2005,over_surface_2012,lundgren_surface_2006,lundgren_kinetic_2004,hendriksen_bistability_2005,hendriksen_looking_2005,hendriksen_role_2010,van_rijn_surface_2011,blomberg_situ_2013}. As a first step to make such systems and phenomena accessible to 1p-kMC we here describe a multi-lattice 1p-kMC approach, which is applicable to morphological transitions between commensurate crystalline lattices. The simple concept is that we work on a super-lattice that contains both lattices simultaneously. In a spatial region describing one stable phase, only the sites of the corresponding lattice are "enabled". The actual morphological transition is then described through the elementary processes of its atomistic pathway, which "disable" sites of one lattice and "enable" sites of the other lattice.

We illustrate this multi-lattice 1p-kMC idea by modeling the surface oxide reduction at Pd(100). Recent experimental work \cite{fernandes_reduction_2014} observed significant changes in the rate of oxide reduction in the temperature range from 300-400\,K, when exposing an initially prepared $(\sqrt{5} \times \sqrt{5})R27^{\circ}$ PdO($101$) surface oxide film (henceforth coined $\sqrt{5}$-oxide for brevity)\cite{todorova_pd1_2003,kostelnik_pd100-55r27o_2007} to a $5\times 10^{-11}\,{\rm bar}$ CO atmosphere. Analysis of the data using fitted mean-field kinetic models suggested that the reduction is controlled by boundary processes between surface oxide and metal domains \cite{fernandes_reduction_2014}. In the present paper, we use extensive DFT calculations to establish an atomistic pathway for the oxide reduction process. Based on this pathway, our multi-lattice 1p-kMC simulations perfectly reproduce the observed trends in the reduction rates, and show that the critical steps in the reduction of the surface oxide correspond indeed to cross-reactions between the metal domains and the surface oxide.

\section{Theory}
\subsection{Kinetic Monte Carlo simulations for heterogeneous catalysis}

We focus on thermally driven surface chemical reactions that involve a site-specific (covalent-type) binding of the reaction intermediates. Such reactions generally follow a so-called {\em rare-event} time evolution. That is, on time scales that can be simulated by {\em ab initio} molecular dynamics all reaction intermediates only vibrate around the lattice sites to which they are bound. Only rarely, i.e. after very many such vibrations, an elementary process changes this population at the lattice sites, where the duration of this process itself is very short. A diffusion process brings a reaction intermediate to another site, adsorption and desorption fill and empty sites at the surface, surface reactions convert reaction intermediates into another etc. During the long time intervals between such individual elementary processes the reaction intermediates equilibrate fully and thereby are assumed to forget through which elementary process they actually reached their current site or state.

Viewing the entire population at all sites as one possible configuration, individual events in form of elementary processes thus advance the system's state from one configuration to another. In each configuration reaction intermediates are assumed to have forgotten their past. In this Markov approximation events will occur uncorrelated from previous ones and the time evolution of the system is described by a Markovian master equation
\begin{equation}
 \dot{\rho}_u(t) = \sum_{v} w_{uv}\rho_{v}(t) - w_{vu}\rho_{u}(t) \quad ,
 \label{eqn:master}
\end{equation}
where $\rho_{u}(t)$ is the probability for the system to be in configuration $u$ at time $t$, and $w_{vu}$ is the transition rate (in units of time$^{-1}$) at which configuration $u$ changes to configuration $v$. For heterogeneous catalysis, solving this equation is in practice not feasible by any direct approach \cite{reuter_first-principles_2012,hoffmann_kmos:_2014}. The kMC approach addresses this problem by generating configuration-to-configuration trajectories in such a way that an average over an ensemble of such trajectories yields the correct probability density $\rho_{u}(t)$ of eq. \ref{eqn:master} \cite{reuter_first-principles_2012,hoffmann_kmos:_2014}. Instead of evaluating and storing the entire transition matrix ${\bf w}$, a kMC code thus only focuses on-the-fly on those transition matrix elements $w_{vu}$ that are actually required to propagate the trajectory.

\subsection{\label{efficient_one_or_more}Efficient 1p-kMC on one or more lattices}

Even with this focus on the required $w_{vu}$, a kMC simulation can still be an extremely demanding task, if the rate constants determining the $w_{vu}$ are to be provided through 1p-calculations. Emerging so-called adaptive or self-learning kMC approaches \cite{kara_off-lattice_2009,trushin_self-learning_2005,xu_adaptive_2008,el-mellouhi_kinetic_2008} address this problem by generating look-up tables of already computed $w_{vu}$ in the course of the simulation. However, an orders of magnitude higher efficiency is needed to make comprehensive 1p-kMC simulations tractable for systems as complex as the present one. This efficiency is achieved when the sites involved in the surface catalysis can be mapped onto a static lattice. Due to the translational symmetry of the lattice, each equivalent site type then features the same group of in principle possible elementary processes, and these elementary processes typically depend only on the occupation of nearby sites. While this does not reduce the number of $w_{vu}$ required during a 1p-kMC trajectory, it dramatically reduces the total number of {\em inequivalent} $w_{vu}$ and therewith 1p rate constants that need to be computed with DFT-TST. In lattice 1p-kMC these 1p rate constants are then in practice precomputed and stored, to be looked up on demand in the evolving kMC simulation.

Assuming a static arrangement of sites, lattice 1p-kMC can not directly be applied to systems undergoing a morphological transition. In the present context, this would be the reduction of the initially oxidized surface to a metallic state, where the oxide overlayer and metal substrate exhibit different crystal lattices. With our suggested generalization to multi-lattice 1p-kMC transitions between commensurable lattices become accessible. The essence of the approach is hereby to choose as basic unit cell of the lattice employed in the kMC simulations a suitable, large enough commensurate cell that embraces all primitive unit cells of the different lattices to be treated. This commensurate cell (and the lattice built up from it) then contains as sites simultaneously all sites of all lattices.

In order to efficiently distinguish (during a simulation) in which lattice state a given local region is currently in, we introduce an additional artificial species (canonically named {\tt null}) to represent "disabled" sites. If say, a certain region is currently in one lattice structure, then the {\tt null} species are placed on all sites corresponding to the other lattices in this region. A site "occupied" with a {\tt null} species is hereby to be distinguished from an empty site. No elementary processes are available at sites occupied by a {\tt null} species, i.e. elementary processes that require an empty site explicitly look for the sites to be empty and not "occupied" by {\tt null}. In a corresponding region where all sites belonging to the other lattices are disabled through {\tt null} species, multi-lattice 1p-kMC thus reduces to a regular lattice 1p-kMC simulation and only the elementary processes corresponding to the 1p-kMC model of this lattice can possibly occur.

In this setup, local transitions between the different lattices are finally enabled by introducing new elementary processes which involve the removal or placement of the {\tt null} species at sites belonging to the lattices involved. Trivially, for the oxide reduction example this could be as simple as an elementary process placing the {\tt null} species on one or more sites of the oxide lattice and removing the {\tt null} species from nearby sites of the metal lattice. As shown below for the Pd oxide reduction, local lattice transitions are unlikely that simple, but follow complex, multi-stepped reaction paths. The additional challenge of setting up and performing a multi-lattice 1p-kMC simulation as compared to traditional lattice 1p-kMC is therefore to establish such atomistic pathways for the lattice transitions and implement them in form of corresponding elementary processes.

From a purely algorithmic point of view as implemented in the kMC modeling framework kmos multi-lattice kMC is identical to single-lattice kMC which underlines that multi-lattice kMC is highly efficient. The only difference is that those elementary steps accounting for morphological changes typically involve significantly more sites (10-20 sites) than typical ordinary surface reaction steps (1-8 sites). The scaling with the number of conditions (i.e. here sites) has been studied in a previous paper\cite{hoffmann_kmos:_2014}. This shows that the computational scaling with the number of conditions is significant, (${\cal O}(n \log(n)$  to ${\cal O}(n^2)$). However in terms of overhead there is no fundamental difference between single-lattice kMC and multi-lattice kMC.

\subsection{\label{sl_models}Literature 1p-kMC models of CO oxidation at the $\sqrt{5}$-surface oxide and at Pd(100)}

Our multi-lattice 1p-kMC simulations describing the reduction of the Pd($100$) surface initially covered with a $\sqrt{5}$-surface oxide film build on existing 1p-kMC models for CO oxidation at the $\sqrt{5}$-oxide and for CO oxidation at Pd($100$). These single-lattice models and the elementary processes considered in them have been established by extensive DFT calculations and have been described in detail before \cite{rogal_co_2008,hoffmann_co_2014}. Here, we only briefly recapitulate their essentials for completeness and as a means to introduce the two lattice types.

\begin{figure}[ht!]
\centering{
\includegraphics[width=7cm]{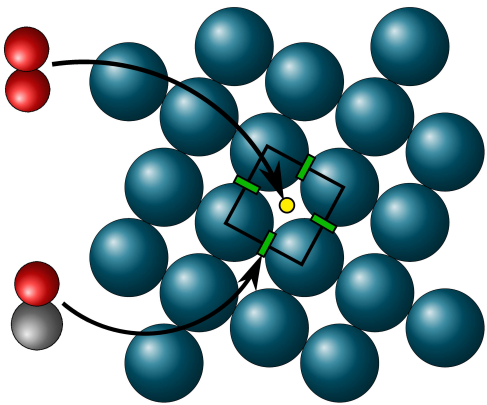}
}
\caption{\label{pd100_model} Illustration of the lattice employed in the 1p-kMC model of CO oxidation at Pd($100$).
 Shown is a top view of the Pd($100$) surface with a primitive unit cell and its hollow (yellow circle) and bridge
 (green rectangle) sites. CO molecules adsorb at bridge sites and oxygen molecules adsorb (dissociatively) at
 hollow sites. Here and henceforth Pd atoms in fcc(100) positions are depicted as large blue spheres, O atoms as small red spheres, and
 C as small grey spheres.}
\end{figure}

Systematic DFT calculations identified the hollow and bridge sites offered by the Pd(100) surface, cf. Fig.~\ref{pd100_model}, as stable high-symmetry adsorption sites for O and CO, respectively \cite{hoffmann_co_2014}. The 1p-kMC CO oxidation model on Pd(100) correspondingly considers all non-concerted adsorption, desorption, diffusion, and Langmuir-Hinshelwood reaction processes involving these sites \cite{hoffmann_co_2014}. Oxygen molecules adsorb dissociatively on two empty next-nearest neighbor hollow sites \cite{brundle_summary_1984,chang_formation_1987,liu_atomistic_2006,liu_atomistic_2009}, CO adsorbs on an empty bridge site. Desorption processes are time-reversals of these non-activated adsorption processes, with rate constants fulfilling detailed balance. Oxygen diffusion occurs between nearest-neighbor hollow sites, CO diffusion between nearest-neighbor bridge sites. Strong short-range repulsive interactions between adsorbed reaction intermediates as calculated by DFT \cite{zhang_accuracy_2007,liu_chemical_2006} are accounted for through site-blocking rules that prevent processes leading to O-O pairs at nearest-neighbor hollow-hollow distances, to CO-CO pairs closer or at next-nearest-neighbor bridge-bridge distance, and O-CO pairs at nearest-neighbor hollow-bridge distance \cite{hoffmann_co_2014}. CO oxidation proceeds via a Langmuir-Hinshelwood mechanism and at the temperatures of interest leads to an instantaneously desorbing CO$_2$ molecule. The rate constants for all these elementary processes have been computed with DFT-TST, specifically employing the PBE exchange-correlation (xc) functional \cite{perdew_generalized_1996}. Typical for CO oxidation at transition metal surfaces, O and CO diffusion barriers are significantly lower than the barriers of all other elementary processes involved. In the 1p-kMC simulations this leads to frequent executions of diffusion events at minute time increments. To increase the numerical efficiency and be able to simulate the extended time scales necessary to follow oxide reduction we artificially increased diffusion barriers for CO of 0.14\,eV and for O of 0.28\,eV\cite{hoffmann_co_2014} by 0.5\,eV. This is permissible, if diffusion is then still fast enough to achieve an equilibration of the adlayer ordering in between the other (rare) elementary processes \cite{reuter_first-principles_2012}. We validated that this is indeed the case, by performing the 1p-kMC simulations with diffusion barriers increased by 0.6\,eV and 0.7\,eV without obtaining any significant changes in the oxide reduction rates reported below.

\begin{figure}[ht!]
 \centering{
 \includegraphics[width=7cm]{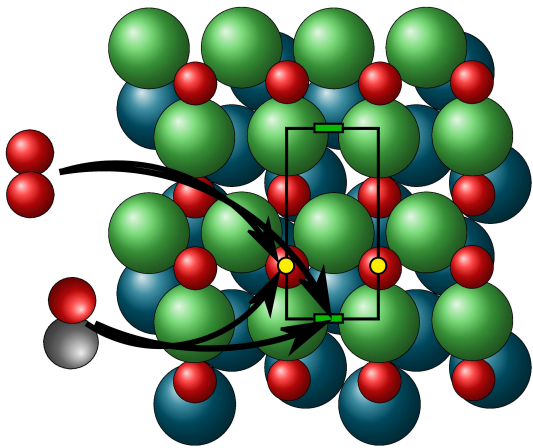}
 }
 \caption{\label{pdsqrt5_model} Illustration of the lattice employed in the 1p-kMC model of CO oxidation at the
 $\sqrt{5}$-oxide. Shown is a top view of the $(\sqrt{5} \times \sqrt{5})R27^{\circ}$ PdO($101$) surface oxide
 reconstruction with a primitive unit cell of the oxide layer and its hollow (yellow circle) and bridge (green rectangle) sites.
 O and CO adsorb at both site types. In the stoichiometric termination shown, O atoms occupy all hollow sites.
 Here and henceforth Pd atoms in surface oxide positions are depicted as large green spheres.}
\end{figure}

As established by detailed first-principles statistical mechanics simulations by Rogal {\em et al.} \cite{rogal_first-principles_2007} only two of the high-symmetry adsorption sites offered by the intact $\sqrt{5}$-oxide are relevant for the reducing environments considered here: a bridge and a hollow site as illustrated in Fig.~\ref{pdsqrt5_model}. Equivalent to the Pd(100) case, the 1p-kMC CO oxidation model at the $\sqrt{5}$-oxide correspondingly considers all non-concerted adsorption, desorption, diffusion, and Langmuir-Hinshelwood reaction processes involving these sites \cite{rogal_co_2008,hoffmann_co_2014}. Eley-Rideal reaction processes as recently suggested by Hirvi {\em et al.} \cite{hirvi_2010} are in principle also considered, but do not play a role at the low CO pressures addressed here. Adsorption is non-activated, apart from dissociative adsorption of O$_2$ into neighboring bridge sites, which is hindered by a sizable adsorption barrier of $1.9$\,eV. As for the Pd($100$) 1p-kMC model, all 1p rate constants of this $\sqrt{5}$-oxide model have been computed with DFT-TST, employing the PBE xc functional \cite{perdew_generalized_1996}. DFT-computed nearest-neighbor lateral interactions are taken into account in these rate constants as detailed by Rogal {\em et al.} \cite{rogal_co_2008}.

\subsection{Computational DFT and kMC settings}

The multi-lattice 1p-kMC simulations of the oxide reduction process consider additional elementary processes to model the atomistic pathway from oxide to metal lattice. This pathway and the involved elementary processes are described in Sections \ref{sec_atomistic_pathway}-\ref{sec_ml_model} below. Their 1p rate constants have been calculated with DFT employing the same TST-based expressions \cite{reuter_first-principles_2006} and the same PBE xc functional \cite{perdew_generalized_1996} as used in the Pd($100$) and $\sqrt{5}$-oxide 1p-kMC mo\-dels. Employing the same computational setup as detailed in ref. \citenum{hoffmann_co_2014}, the DFT calculations have been performed with the {\tt CASTEP} package \cite{clark_first_2005} via the {\tt Atom\-ic Simulation Environment} \cite{bahn_object-oriented_2002}. Previous work \cite{hoffmann_co_2014} established that the employed standard library ultrasoft pseudopotentials accurately reproduce the full-potential $\sqrt{5}$-oxide adsorption energetics of Rogal {\em et al.} \cite{rogal_co_2008}.
Geometry optimizations and tran\-si\-tion-state searches were done with supercell geometries containing asymmetric slabs with one (frozen) layer of Pd($100$) and one layer of PdO($101$), separated by at least $10$\,{\AA} vacuum. Geometry optimizations were performed using the BFGS method with a force threshold of $0.05$\,eV/{\AA}. The energetics of the thus obtained intermediate geometries were refined in single-point calculations with further Pd(100) layers added to the slab. Transition state searches have been performed with the climbing-image nudged elastic band method \cite{henkelman_climbing_2000} or simple drag procedures with the nature of the transition state confirmed through frequency calculations. From test calculations with additional Pd(100) layers added we estimate the uncertainty in the thus determined barrier values as 0.1 eV. The gas-phase calculations for O$_2$ and CO were performed in a cubic supercell of $15$\,{\AA} box length and $\Gamma$-point sampling. At a plane-wave cutoff of $400$\,eV and a  [$4\times 4\times 1$] $k$-point grid for the commensurate surface oxide unit cell and a [$2\times 2\times 1$] $k$-point grid  for the ($2\times 2$) surface oxide unit cell in the slab calculations convergence tests demonstrate a numerical convergence of the obtained DFT barriers and binding energies to within 0.1\,eV.

All 1p-kMC simulations have been performed using the {\tt kmos} framework \cite{hoffmann_kmos:_2014}. The periodic boundary simulation cell contains $(40 \times 20)$ $\sqrt{5}$-unit cells and thus 1600 sites in case of the single-lattice 1p-kMC simulations on the surface oxide. Systematic tests with varying simulation cell sizes showed the obtained oxide reduction rates to be fully converged at this setting. In the multi-lattice case one commensurate unit-cell contains already a total of 19 sites from both the metal and the oxide lattice. We employed periodic boundary simulation cells comprising $(10 \times 20)$, $(20 \times 20)$ and $(40 \times 20)$ such commensurate unit-cells to specifically assess the role of elementary processes across oxide and metal domains (see below). In all cases, averages over 10 kMC trajectories starting with different random number seeds were taken to obtain the correct transient quantities.

\section{Results}

\subsection{\label{singlat_sim}Single-lattice 1p-kMC simulations of oxide reduction}

As mentioned initially the central objective of our simulations is to analyze the oxide reduction rates obtained in recent high-resolution X-ray photoelectron spec\-tros\-co\-py experiments by Fernandes\,{\em et al.} \cite{fernandes_reduction_2014}. Within a small range of temperatures, they reported large variations for the total time required to fully reduce an initially prepared $\sqrt{5}$-oxide layer by exposure to a pure CO atmosphere of $5\times10^{-11}\,{\rm bar}$. Specifically, at 303\,K the oxide was reduced after approximately 90\,minutes, while the same process took only approximately 12\,min at 343\,K and less than 4\,min at 393\,K. Previous theoretical work addressing the surface oxide stability employed the 1p-kMC approach with a single lattice, namely the $\sqrt{5}$-oxide 1p-kMC model. They monitored the coverage of O in the hollow sites as an indicator for the oxide stability \cite{rogal_co_2008}. Using this criterion 1p-kMC simulations fail qualitatively to reproduce the experimental findings of Fernandes\,{\em et al.}. Even at the highest temperature of 393\,K no reduction of the oxide is observed at all after 120\,mins, i.e. the oxygen coverage remains essentially intact over the entire time. This can be explained by the lack of CO on the surface, which implies that an efficient removal of oxygen via CO oxidation is not possible.

\begin{figure}
 (a)\includegraphics[width=8cm]{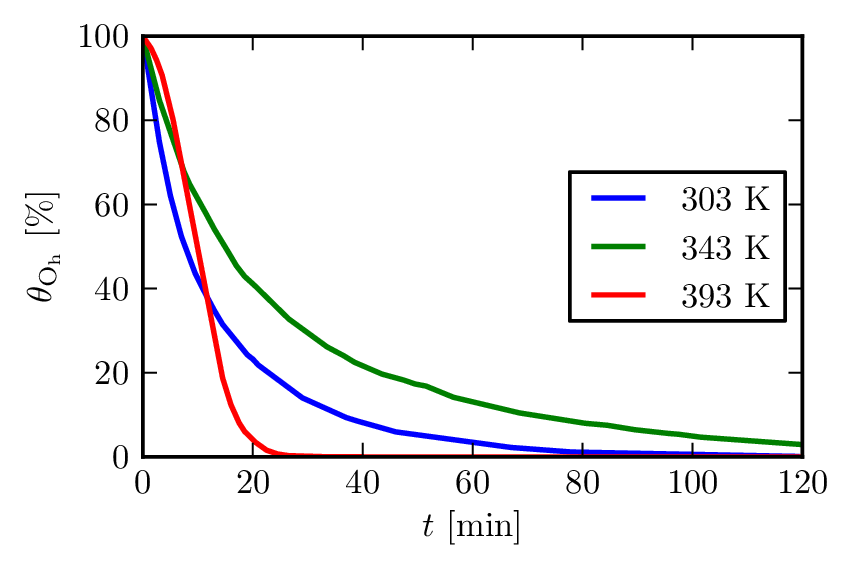}
 (b)\includegraphics[width=8cm]{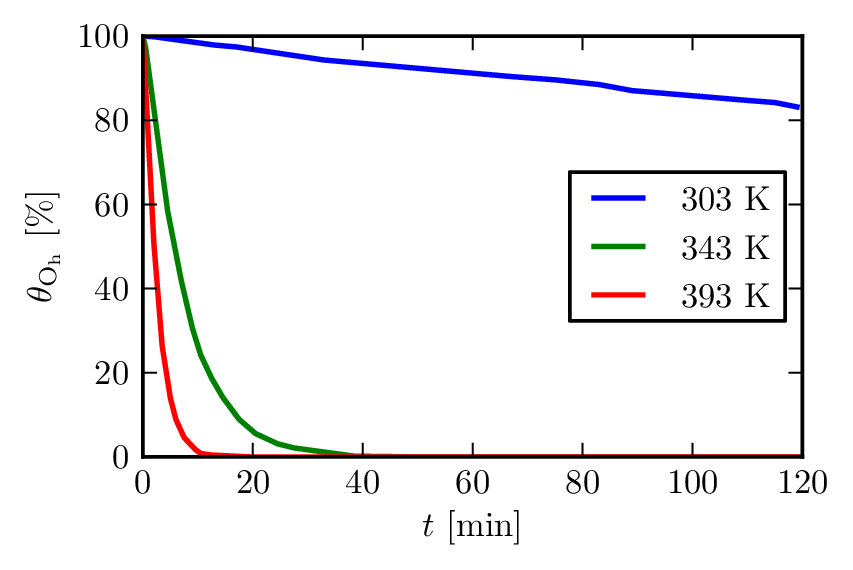}
 \caption{\label{slkmc} Single-lattice kMC simulations for ${\sqrt 5}$-oxide
 reduction at the three temperatures employed in the experiments and using the O coverage at the
 hollow sites ($\theta_{\rm O_h}$) as indicator for the progressing reduction of the
 surface oxide (see text). The 1p rate constants are modified to achieve at least
 some agreement with the experimental findings: (a) CO adsorption energy lowered by
 0.2\,eV. The reduction times for all three temperatures are in the right order of magnitude,
 but the ordering with temperature is wrong. (b) CO adsorption energy lowered by 0.5\,eV;
 CO oxidation barrier increased by 0.15\,eV. The oxide reduction time correctly decreases
 with temperature, but the reduction time at 303\,K disagrees qualitatively with experiment.}
\end{figure}

Since an error in the reported experimental conditions large enough to explain this discrepancy seems unlikely we focus on the uncertainty in the DFT parameters deployed in the model.
A possible reason for this discrepancy could be that the relevant CO adsorption energy of $-0.92$\,eV as calculated by DFT-PBE is too small, and/or the CO reaction barrier towards CO$_2$ formation of 0.9\,eV\cite{rogal_co_2008} is too high. We explore these possibilities by running the 1p-kMC simulations using modified rate constants. Specifically, we vary the CO oxidation barrier and the CO adsorption energy on the bridge and hollow sites within $\pm 1$\,eV. However, no such kMC simulation reproduces the experimentally observed behavior: Either the time scale for reduction is too long or too short for all three temperatures, or if the simulations exhibit a correct time scale for at least one of the temperatures, then the trend with temperature is wrong, i.e. the reduction time does not decrease for each higher temperature as found in the measurements. We illustrate this in Fig.~\ref{slkmc} for two cases where at least partial agreement with experiment is obtained: When the CO adsorption energy is lowered by 0.2\,eV and the CO reaction barrier is left unchanged, reasonable reduction time scales are obtained for all three temperatures, but the ordering is wrong: At the intermediate temperature of 343\,K reduction is slower than at the lower temperature of 303\,K. When instead the CO adsorption energy is lowered by 0.5\,eV and the CO oxidation barrier is simultaneously increased by 0.15\,eV, this ordering is correct, but the time scale for oxide reduction at 303\,K is much too long.

It is difficult to generally rule out that also varying the rate constants of all other elementary processes would not coincidentally yield a parameter set that reproduces the experimental reduction times. Nevertheless, the failure to obtain such a parameter set when modifying the two processes that most obviously relate to the reduction time, let us believe that a shortcoming of the employed 1p rate constants is unlikely the reason for the qualitative discrepancy to the measurements. We therefore tend to conclude that the reduction of the $\sqrt{5}$-oxide layer does not happen at intact oxide patches (for which the O coverage at the hollow sites is a useful stability indicator).

\subsection{Atomistic pathway for the $\sqrt{5}$-oxide reduction\label{sec_atomistic_pathway}}

\begin{figure}
 \centering{
 \includegraphics[width=7cm]{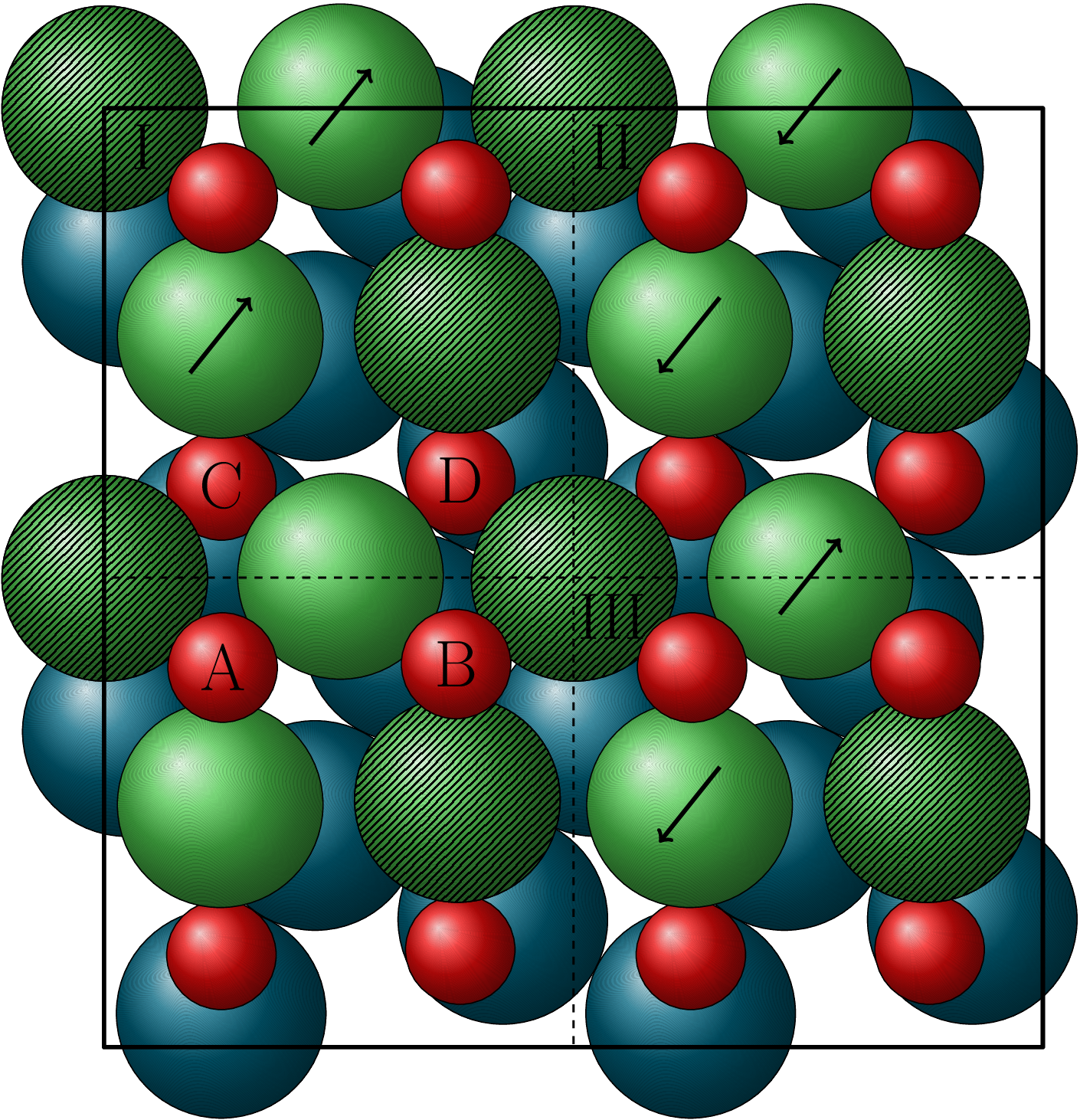}
 }
 \caption{\label{reconstruction_considerations}
 Top view of the supercell model employed in the oxide reduction DFT calculations, spanning
 four $\sqrt{5}$-oxide unit cells in a $(2 \times 2)$ arrangement. Each intact $\sqrt{5}$-oxide unit
 cell contains four O atoms, labeled from A to D, and four Pd atoms, two of which (marked through
 hatching) are located in hollow sites of the underlying Pd($100$) substrate. The arrows in the upper
 right three unit-cells indicate the directions in which the two Pd atoms can be shifted to generate
 a minimal Pd($100$) patch: This defines the three cases I (upper left unit-cell), II (upper right
 unit-cell) and III (lower right unit-cell) referred to in the text.}
\end{figure}

In the context of oxide formation and reduction efficient cross-reactions at the boundary between oxide and metal domains have long been discussed \cite{imbihl_1995}. Fernandes {\em et al.} reported long induction times of the order of minutes at all measured temperatures \cite{fernandes_reduction_2014}. This renders a crucial role of cross-reactions at extended surface imperfections like steps or dislocations rather unlikely. Instead we suspect that a slow initial reduction of ideal oxide terraces locally generates minimal Pd(100) patches that then enable an efficient continued reduction through cross-reactions. Admittedly, also for this suggestion the observed induction time is somewhat long. In order to scrutinize this picture with multi-lattice 1p-kMC simulations we therefore establish an atomistic pathway from the intact $\sqrt{5}$-oxide to reduced Pd($100$) and focus on the structural evolution of the surface oxide film when progressively removing O atoms. These studies are consistently performed with large supercell models spanning four $\sqrt{5}$-oxide surface unit cells in a $(2 \times 2)$ arrangement as shown in Fig.~\ref{reconstruction_considerations}. This model is large enough to study the effects of O removal on one $\sqrt{5}$ cell that is still surrounded by intact $\sqrt{5}$-oxide units in every direction. For these studies, an important structural aspect of the commensurate $\sqrt{5}$-oxide/Pd($100$) interface is that in every $\sqrt{5}$-oxide unit cell two of the four Pd atoms in the oxide layer are located in fourfold hollow sites of the underlying Pd($100$) substrate, with the other two being rather close to bridge sites, cf. Fig.~\ref{reconstruction_considerations}. After complete reduction, the Pd atoms of the surface oxide will form domains of a new Pd($100$) layer \cite{todorova_pd1_2003,lundgren_private_2014}, in which they are also located in the fourfold hollow sites. This means that during the local oxide decomposition two out of four Pd atoms in the oxide layer do not necessarily have to significantly change their lateral position.

In the stoichiometric $\sqrt{5}$-oxide termination there are four O atoms per $\sqrt{5}$ unit cell. Two are located above the Pd atoms of the oxide layer (occupying the hollow sites described above) and two below, as shown in Fig.~\ref{reconstruction_considerations}. In agreement to previous DFT studies performed in smaller supercells \cite{rogal_first-principles_2007}, we find no significant structural changes upon removal of any single of these four O atoms in one $\sqrt{5}$ unit cell of our model. In particular, none of the oxide Pd atoms change their lateral registry. The same is observed, when removing pairs of O atoms in all possible combinations. Apparently neither single O vacancies, nor O divacancies lead to an unactivated destabilization of the $\sqrt{5}$-oxide structure.

\begin{figure*}[ht]
 \centering{
 (a)\includegraphics[width=4.5cm]{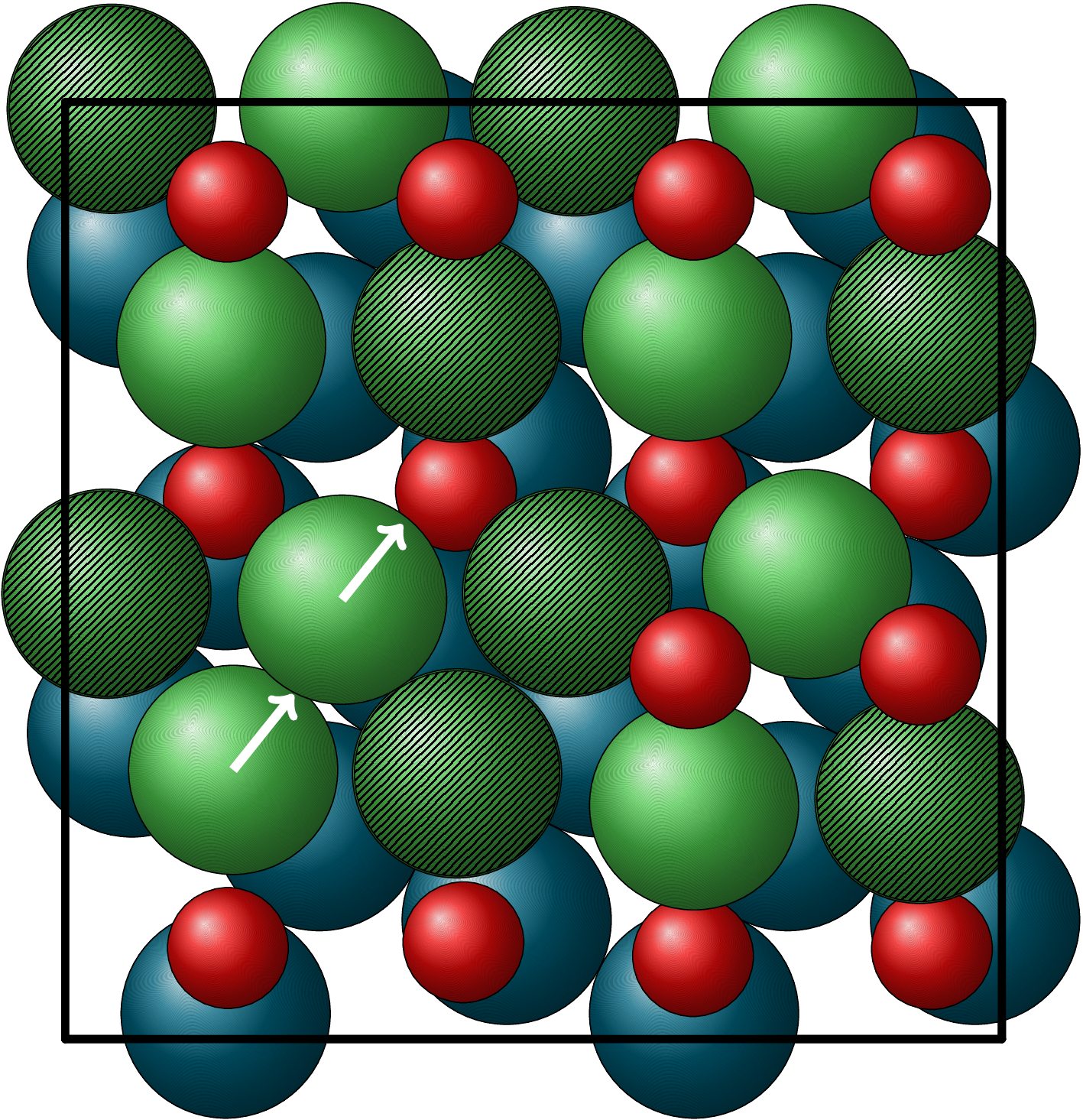}
 (b)\includegraphics[width=4.5cm]{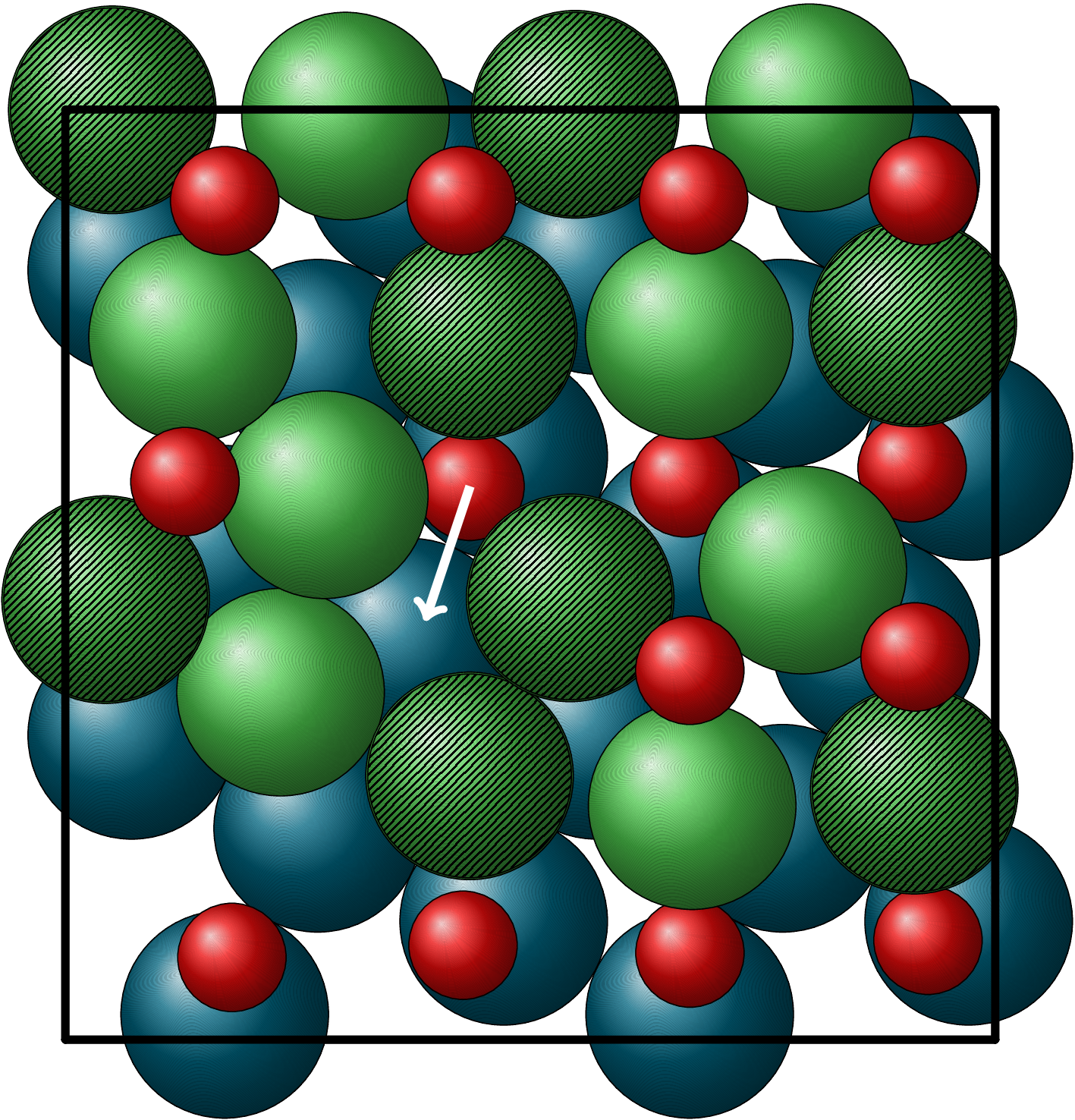}
 (c)\includegraphics[width=4.5cm]{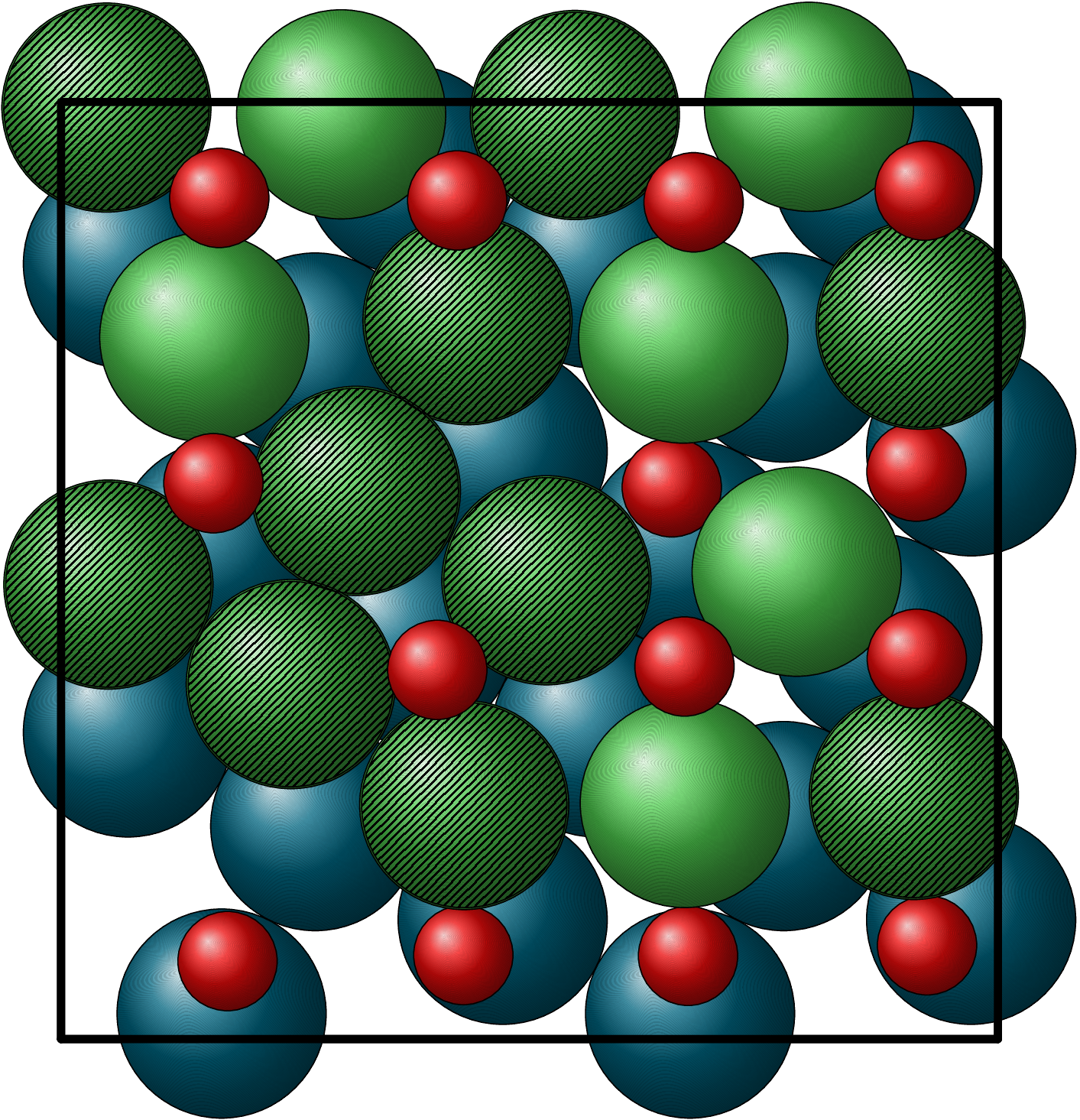}
 }
		\caption{\label{intermediate_structures}
		Top view of intermediate structures along the suggested oxide reduction pathway. (a) Divacancy
		created by removal of two neighboring upper O atoms in one $\sqrt{5}$-oxide unit-cell (labeled A and B
		in Fig. \ref{reconstruction_considerations}). (b) Intermediate structure formed from (a) through an
		essentially unactivated shift of the two bridge-site Pd atoms as in case I of
		Fig.~\ref{reconstruction_considerations}. This structure is by 0.4\,eV more stable than structure (a).
		(c) Pd($100$) patch formed from (b) after the activated diffusion of the sub-surface O labeled D in
		Fig. \ref{reconstruction_considerations} to the fourfold on-surface site in the center of
		the thus formed Pd($100$) patch (see text).}
\end{figure*}

In this situation, we focus on a straightforward way to generate a minimal Pd($100$) patch within the surface oxide layer. We explore if this (together with an increasing number of O vacancies) does not constitute a pathway with at least a very low activation barrier. This pathway centers on the above mentioned realization that two of the four Pd atoms in the $\sqrt{5}$-oxide unit cell are already located in fourfold hollow sites with respect to the underlying Pd($100$) substrate. A direct way to convert one entire $\sqrt{5}$ cell into a Pd($100$)-type patch would therefore be to simply shift the other two Pd atoms from their close to bridge positions to adjacent hollow sites. As illustrated in Fig.~\ref{reconstruction_considerations} there are three ways to realize this: Either both Pd atoms move to hollow sites in the same direction (cases I and II), or the two Pd atoms move to hollow sites in opposite directions (case III). If we enforce such lateral shifts on the Pd atoms of one unit cell for the intact surface oxide, they relax back to their original positions upon geometry optimization. The same holds, if this is done with one O vacancy in the $\sqrt{5}$-oxide unit cell, or when moving only one of the two Pd atoms.

In contrast, metastable structures result when the same movements are applied in the presence of an O divacancy. Under reaction conditions such an O divacancy could arise from the random reduction of neighboring O atoms in the oxide but possibly also due to a consequential reaction step next to a metal to surface oxide boundary. Following a systematic development from the fully intact surface oxide similar to a cluster expansion divacancies are an obvious next candidate motif after single vacancies. Specifically this is when removing the two upper O atoms, labeled A and B in Fig.~\ref{reconstruction_considerations}, that are directly coordinated to one of the moving Pd atoms. For the three cases defined in Fig.~\ref{reconstruction_considerations} we calculate the relative energies of the resulting metastable structures compared to the original divacancy structure with the Pd atoms in their regular positions. They are $-0.40$\,eV (I), $-0.25$\,eV (II) and $+0.17$\,eV (III). Cases I and II lead therefore to more stable intermediate structures. In particular for the most stable case I we furthermore find only a negligibly small activation barrier below 0.1\,eV.

This suggests the following initial steps in the oxide reduction pathway: Upon local formation of an O divacancy as shown in Fig.~ \ref{intermediate_structures}a, the two bridge-site Pd atoms of the corresponding $\sqrt{5}$-oxide cell shift laterally and essentially unactivated to yield the intermediate structure shown in Fig.~\ref{intermediate_structures}b. This metastable structure is more stable than the divacancy by 0.4\,eV. It already features a minimal Pd($100$) patch with four Pd atoms in hollow sites, but still contains the two O atoms of the $\sqrt{5}$-oxide cell that were originally below the Pd layer. One of these two, labeled C in Fig.~\ref{reconstruction_considerations}, automatically moves to a position above the Pd atoms of the Pd($100$) patch along with the lateral shift of the bridge-site Pd atoms. The other O atom D remains in a sub-surface position as depicted in Fig.~\ref{intermediate_structures}b.

We calculate a barrier of 0.66\,eV for the diffusion process that lets this O atom pop up to the on-surface site E in the center of the Pd($100$) patch, cf. Fig.~\ref{intermediate_structures}c. There the O atom is by 0.11\,eV more stable than in its original sub-surface site. From there and as further detailed below, the on-surface O atom can be reacted off through Langmuir-Hinshelwood-type CO oxidation. Alternatively, it can diffuse away. Both cases leave a bare minimal Pd($100$) patch as the end product of the here suggested oxide reduction pathway. Continued reduction of the oxide layer will then lead to an increasing number of such Pd($100$) patches. As the Pd density in the intact surface oxide is 20\,\% lower than in a Pd($100$) layer, Pd vacancies will emerge along with the formation of these patches. The diffusion of such Pd vacancies will be very fast. The calculated activation barrier of only $\sim 0.2$\,eV is similar to the diffusion of Pd on Pd($100$) \cite{liu_eam_1991,evans_surface_1993,kim_transition-pathway_2007,kim_transition-pathway_2007-1,liu_atomic_2014}. Quickly, this fast diffusion will lead to the coalescence of the individual Pd($100$) patches and with that to the formation of extended Pd($100$) islands as seen in experiment \cite{evans_surface_1993,hendriksen_bistability_2005,hendriksen_looking_2005,hendriksen_role_2010}.

\subsection{Elementary processes at the Pd($100$) patch under reduction conditions\label{sec_elementary_processes}}

\begin{figure}[ht]
 \centering{
 \includegraphics[width=7cm]{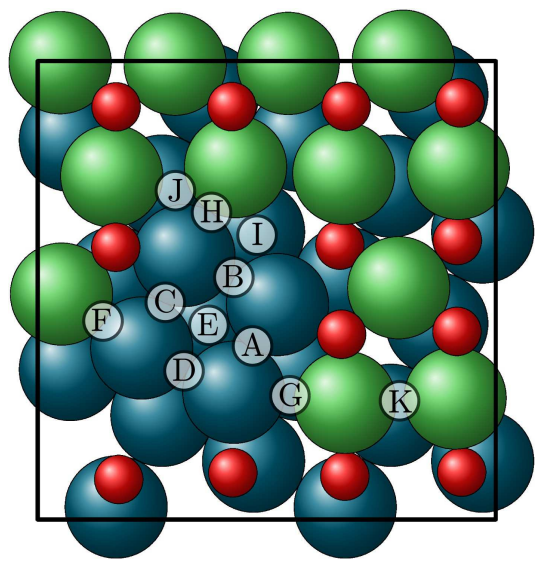}
 }
 \caption{\label{defect_sites} Top view of the Pd($100$) patch as formed during the oxide reduction
		process. Shown are all high-symmetry binding sites explored to assess the binding properties of the
		patch itself (sites A-E) and at its boundary (sites F-K). Pd atoms of the formed patch are darkened.}
\end{figure}

The Pd($100$) patch is the central new feature within the obtained oxide reduction pathway. During initial oxide reduction it may emerge upon sufficient local oxygen depletion of the $\sqrt{5}$-oxide. Its existence may then enable new elementary processes specific to either the patch itself or the boundary between patch and surrounding oxide. As the next step we therefore proceed to a systematic identification and description of corresponding processes. As starting point for this we first consider the binding at the different high-symmetry sites illustrated in Fig. \ref{defect_sites}. This comprises sites on top of the patch and at its boundary. Specifically, sites labeled A-D and site E correspond to Pd($100$)-type bridge and hollow sites on the patch, respectively. Sites F-J are sites directly at the patch/oxide boundary, whereas site K corresponds to a $\sqrt{5}$-oxide bridge site in the immediate vicinity of the patch. Table \ref{defect_energies} summarizes the computed binding energies at these sites for the two "oxidation" stages of the patch along the oxide reduction pathway: With the sub-surface oxygen atom still beneath it, cf. Fig. \ref{intermediate_structures}b, and without it. The latter situation results as a consequence of the activated diffusion of the O atom to the on-surface hollow site of the patch and subsequent reaction or diffusion, {\em vide infra}, and is the situation depicted in Fig. \ref{defect_sites}.

\begin{table}
\caption{\label{defect_energies}
Calculated DFT binding energies for CO and O at high-symmetry sites on top of the Pd($100$) patch and at its boundary. Shown are results for the situation with the sub-surface O atom still beneath the patch (labeled "w/ sub-surf O") and without the sub-surface O atom (labeled "w/o sub-surf O"). Sites A-K are defined in Fig. \ref{defect_sites}. Fields marked with a letter indicate that the adsorbate moved to the respective site during geometry optimization. Fields marked "$-$" indicate sites for which no binding was obtained. Binding in sites marked with an asterisk ended up in a threefold coordination after geometry optimization. All energies are in eV.}
\centering{
\begin{tabular}[t]{l|r r|r r}
  & \multicolumn{2}{c|}{w/ sub-surf O} &\multicolumn{2}{c}{w/o sub-surf O} \\ \hline
  & {\rm CO} & {\rm O }& {\rm CO}     & {\rm O } \\ \hline
A & -1.72    &    E    &   -1.85      &    E     \\
B &  $-$     &   $-$   &   -1.73      &    I     \\
C & -1.40    &    D    &   -1.62      & -0.98$^{*}$ \\
D & -2.06    &  -0.96  &   -2.06      & -1.00    \\ \hline
E &   D      &  -1.22  &   -1.90$^{*}$& -1.25    \\ \hline
F & -1.49    &  -0.96  &   -1.72      & -1.25    \\
G & -1.27    &  -0.40  &   -1.31      & -0.47    \\
H & -0.83    &  J      &   -1.56      & J        \\
I & $-$      &  $-$    &   B          & -0.96    \\
J & H        &  -0.39  &   H          & -0.75    \\ \hline
K & -1.03    & -0.30   &   -1.10      & -0.37    \\
\end{tabular}
}
\end{table}

With the exception of site B directly adjacent to the sub-surface O atom, the effect of the latter atom on the binding properties of the sites on top of the patch is rather small. Quite similar binding energies are obtained for these sites for the two cases. {\em Cum grano salis} these binding energies on the patch sites are also quite similar to those computed for the extended Pd($100$) surface. Roughly we obtain a preferred O binding at the hollow site E of about $-1.25$\,eV. This should be compared to $-1.25$\,eV at the corresponding site on Pd($100$). In turn, the calculated preferred CO binding at the bridge sites A-D of about $\sim$ $-1.4$-$2.1$\,eV should be compared to $-1.93$\,eV at bridge sites of Pd($100$). Without the presence of the sub-surface O atom this also extends to the bridge sites F and H at the interface between the patch and the $\sqrt{5}$-oxide. Again the CO binding properties are very similar to regular bridge sites at extended Pd($100$). Furthermore considering the close geometric and energetic vicinity of these different sites we therefore also expect similarly low diffusion barriers as on Pd(100). Correspondingly, there will be a predominant presence of the adsorbates in the most favorable sites, i.e. CO in the bridge site D and O in the hollow site E at the temperatures employed in the oxide reduction experiments.

Multiple adsorption on one patch in form of CO adsorption at several bridge sites, or O and CO coadsorption on hollow and bridge site, respectively, is not possible according to our DFT calculations. Also in this respect, adsorption at the Pd(100) patch seems very similar to adsorption at extended Pd(100). Likewise the disturbance of the $\sqrt{5}$-oxide created by the Pd(100) patch seems quite short-ranged. Already at the oxide bridge site K directly adjacent to the patch, O and CO binding is essentially identical to the binding at the bridge site on the stoichiometric $\sqrt{5}$ termination. The CO binding is calculated as $-1.03$\,eV and $-1.10$\,eV at site K (with and without sub-surface O underneath the adjacent Pd(100) patch) and as $-1.08$\,eV on the intact $\sqrt{5}$-oxide. The O binding energy is $0.30$\,eV, $-0.37$\,eV and $-0.19$\,eV for the three cases, respectively. Finally, the binding at the interfacial sites F, G, H and I in the presence of the sub-surface oxygen atom and at the interfacial site H without the sub-surface O atom is intermediate to the one obtained on the patch and on the intact $\sqrt{5}$-oxide.

In light of these results, we can now systematically analyze the new elementary processes that may arise due to the patch under the reducing conditions of the experiment and assess their qualitative relevance for the oxide reduction process. CO adsorption and desorption processes can take place at the patch bridge sites, with a maximum coverage of one CO molecule on a patch. This is regardless of whether the sub-surface O atom is still present underneath the patch or not. CO diffusion on the patch itself will be fast, and will lead to a preferential population of bridge site D at the temperatures employed in the reduction experiments. After coalescence of nuclei to larger Pd(100) domains, equivalent processes (CO adsorption, desorption, diffusion) will occur, with properties as on extended Pd(100). In principle, CO diffusion could also occur from patch (or Pd(100) domain) to oxide or vice versa. An important point to realize here, however, is that CO binding at the patch (and at Pd(100)) is about 1\,eV more stable than at the hollow or bridge sites of the $\sqrt{5}$-oxide. This already imposes a quite large mere thermodynamic barrier for CO diffusion onto the oxide. Simultaneously CO diffusion processes from the oxide to the patch will be rare, owing to the near zero CO coverage on the surface oxide under the conditions of the reduction experiments, cf. Section~\ref{singlat_sim}.

If all sites A-E of the patch are unoccupied, the sub-surface O atom can pop up into the on-surface fourfold hollow site E of the patch, with a calculated barrier of 0.66\,eV. The reverse process is possible with a corresponding barrier of 0.77\,eV. From the patch site E the oxygen atom can diffuse to an adjacent vacant hollow site of the surrounding oxide, or vice versa. Since O binding is about 0.4\,eV more stable at the oxide hollow sites, this diffusion will preferentially take oxygen away from the patch (or coalesced larger Pd(100) domains). For O diffusion between the oxide hollow sites directly adjacent to the patch (those above site K in Fig. \ref{defect_sites}) we find a slightly lowered diffusion barrier of 1.1\,eV as compared to the 1.4\,eV for the diffusion over the intact surface oxide. This lowering can be rationalized with the higher geometric flexibility of the underlying Pd atoms in the partially reduced surface oxide environment. Oxygen diffusion from site E to adjacent bridge sites of the $\sqrt{5}$-oxide face a similarly large thermodynamic barrier as the CO diffusion processes due to the much weaker O binding at these oxide sites. As these bridge sites are furthermore essentially all unoccupied under the conditions of the reduction experiments, the reverse diffusion processes bringing O atoms from adjacent oxide bridge sites to the patch will also occur at a negligibly low rate. Associative oxygen desorption requires two nearby O atoms. It can thus only take place either after coalescence of nuclei (which creates next-nerest neighbor Pd(100)-type hollow sites) or in case of an individual patch by involving one O atom from the patch site E and one O atom from nearest neighbor O hollow sites of the oxide. Due to the similar binding energies of O on both metal and oxide, such processes are as rare as they were in the single-lattice 1p-kMC simulations focusing only on the surface oxide. Correspondingly, such processes are unlikely to have a relevant effect on the reduction kinetics.

\begin{figure}
 \centering{
 \includegraphics[width=7cm]{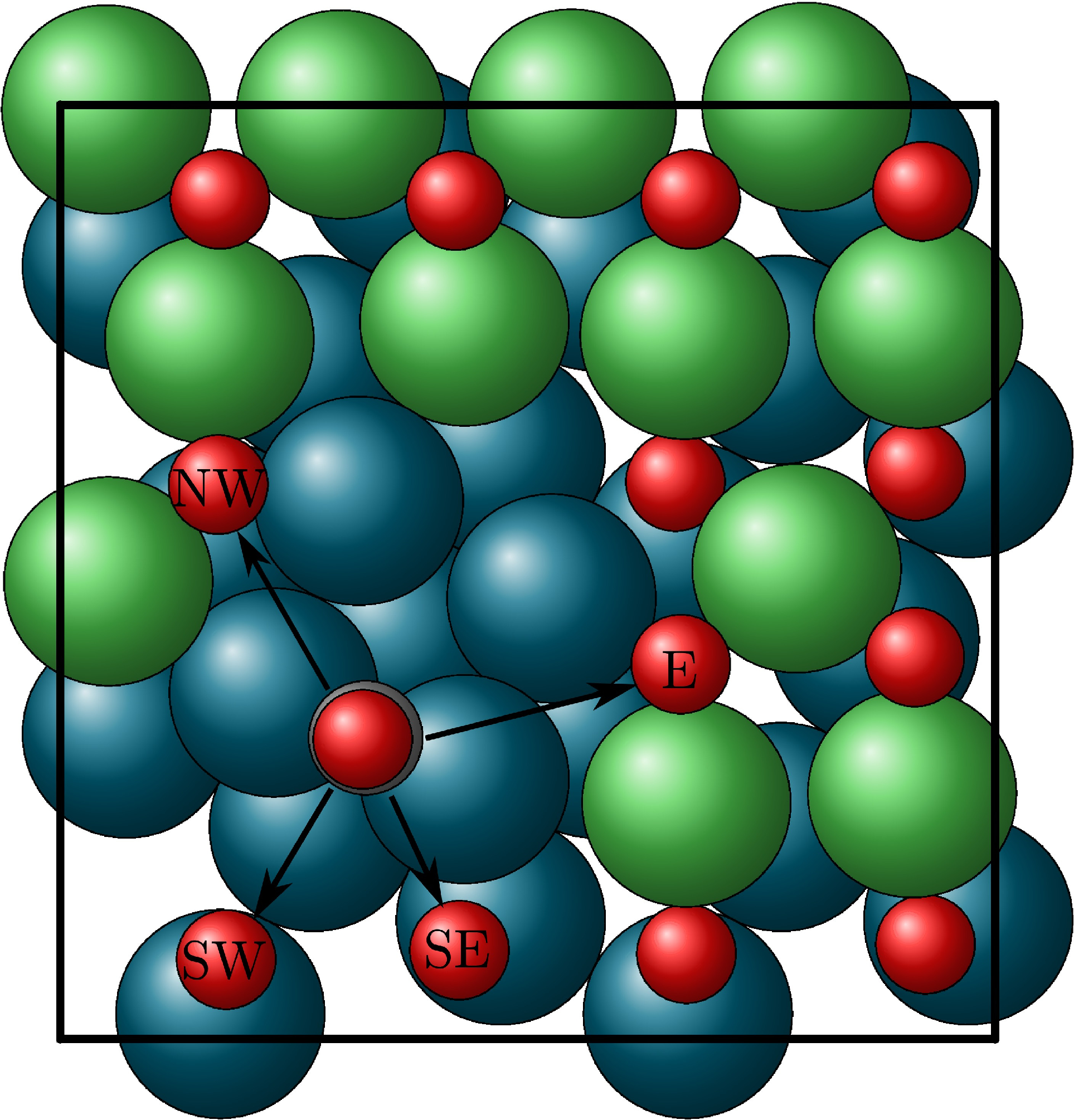}
 }
 \caption{\label{reaction_partners_CO_D} Top view of the Pd(100) patch illustrating possible CO
	 oxidation reactions across the patch/oxide boundary. The CO molecule is adsorbed at the most
	 stable bridge site D of the Pd(100)-patch. The calculated barriers for the four reaction
	 processes are: NW $1.63$\,eV, SW $1.46$\,eV, SE $1.54$\,eV, and E $0.95$\,eV.}
\end{figure}

Finally, new CO oxidation reactions can be enabled at the patch or its boundary. One possibility is a reaction of a CO molecule adsorbed on the patch with the sub-surface O atom. For a CO molecule occupying the preferred bridge site D we calculate a barrier of 1.27\,eV for this process. This is rather high for this process to play a significant role at the temperatures employed in the experiments. Alternatively, the CO molecule could react with O atoms of the surrounding $\sqrt{5}$-oxide. Focusing on the preferred binding site D Fig. \ref{reaction_partners_CO_D} illustrates the four corresponding reactions to the closest such O atoms and compiles the calculated barriers for these processes. Only one of these processes exhibits a barrier below 1\,eV and should therewith exhibit a similar reactivity as the reaction processes occurring on either Pd(100) or on the $\sqrt{5}$-oxide. As last possibility, an O atom occupying the on-surface site E of the patch could react with a nearby adsorbed CO molecule. Due to the near zero CO coverage on the oxide during the reduction experiments, such a process is unlikely to occur with a CO molecule adsorbed on the $\sqrt{5}$ though. At the same time, the strong repulsive interactions prevent CO adsorption at the bridge sites A-D of the patch, if an O atom occupies the hollow site E. This leaves as only possibility for such a cross-reaction, the reaction with a CO molecule adsorbed at the bridge site F, cf. Fig. \ref{defect_sites}. In general, such a reaction process can not have a strong influence on the oxide reduction process though. As O diffusion from the $\sqrt{5}$-oxide onto the patch is rare, this process will predominantly occur after the sub-surface O atom has popped up to the surface. The process then takes place once per formed patch and is thereafter unlikely to reoccur.

Summarizing these insights, the qualitatively new feature emerging from the presence of a formed Pd(100) patch during the oxide reduction process is the significantly enhanced stabilization of CO at the patch as compared to the $\sqrt{5}$-oxide. Whereas the CO coverage on the latter is close to zero under the reducing conditions of the experiment, a much higher CO coverage can thus be expected at formed patches. An oxidation reaction of such CO molecules across the patch/oxide interface with O atoms of the surface oxide then provides a much more efficient means to deplete the surface oxygen compared to the two mechanisms possible at the $\sqrt{5}$-oxide alone. There the associative O desorption is a rare process due to the strong O binding energy on the oxide. Similarly the CO oxidation is a rare process due to the low CO coverage at the surface. In contrast, the new reduction process enabled at the oxide/patch interface only depends on the CO adsorption, which is comparatively frequent in the reducing environment of the experiment, and on the replenishment of the created O vacancies at the oxide sites at the interface. The latter occurs via O diffusion on the surface oxide, which is not a very fast process. Yet, it is still faster than the oxide-only reaction mechanism, considering in particular that the number of formed Pd(100) patches will scale with the increasing number of concomitantly created O vacancies in the oxide layer.

\subsection{Multi-lattice 1p-kMC model \label{sec_ml_model}}

We now proceed by casting the established picture of the reduction process via the formation of Pd(100) patches into an explicit multi-lattice 1p-kMC model. For regions of Pd(100) and intact $\sqrt{5}$-oxide this multi-lattice 1p-kMC model will employ the literature single-lattice 1p-kMC models as before. The new elementary processes arising in the context of the Pd(100) patches and their DFT calculated barriers are listed as follows:

A local reduction begins whenever an oxygen divacancy involving two neighboring hollow sites on the $\sqrt{5}$-oxide, cf. Fig. \ref{intermediate_structures}, is formed during the 1p-kMC simulation. The patch formation is then modeled as a non-activated process, i.e. it automatically follows the formation of the divacancy. In terms of the multi-lattice representation the patch formation corresponds to deactivating the two bridge and two hollow sites associated with the corresponding $\sqrt{5}$-oxide unit cell by changing their occupation status to {\tt null}. In their place, Pd(100) lattice sites are activated. Specifically, this concerns Pd(100) bridge sites at sites labeled A, B, C, D, F, H in Fig.~\ref{defect_sites}, the occupation of which is changed from {\tt null} to {\tt empty}. Furthermore, a Pd($100$) hollow site is created {\tt empty} at the site labeled E in Fig.~\ref{defect_sites}. The quenched-in oxygen atom is modeled to remain at its sub-surface $\sqrt{5}$-oxide site. However, now that the Pd($100$) hollow site labeled E in Fig.~\ref{defect_sites} is created, this sub-surface O atom can hop there with a barrier of 0.66\,eV. However, this only, if the Pd(100) hollow site and its four nearest-neighbor Pd(100) bridge sites are empty in accordance with the site-blocking rules used in the Pd(100) single-lattice 1p-kMC model \cite{hoffmann_co_2014}. From there, the oxygen atom can reverse the diffusion with a barrier of 0.77\,eV. Alternatively, if there are already other immediately adjacent Pd(100) patches, it can also regularly diffuse to corresponding nearest-neighbor Pd(100) hollow sites. O diffusion from the patch to surrounding $\sqrt{5}$-oxide (and the reverse back-diffusion process) exhibits a barrier of 0.5\,eV (0.9\,eV). The hollow-hollow diffusion of oxygen on the $\sqrt{5}$-oxide immediately adjacent to a Pd(100)-patch has a barrier of 1.1\,eV, and is therewith accelerated as compared to diffusion on the intact $\sqrt{5}$ with its barrier of 1.4\,eV.

The second oxygen atom that is originally in the sub-surface $\sqrt{5}$-oxide site and pops up to the surface during the formation of the Pd(100) patch, cf. Fig. \ref{intermediate_structures}b, has a binding energy of $-1.42$\,eV, irrespective of whether the other adjacent quenched-in sub-surface O atom is still present or not. Even though thus exhibiting a somewhat stronger binding by 0.17\,eV, we still approximate this as a regular O atom in a Pd(100) hollow site.  In the multi-lattice scheme we correspondingly activate another Pd(100) hollow lattice site at the location that is obtained by going from site E one lattice constant to the left in Fig.~\ref{defect_sites} and directly occupy it with the popped-up O atom. Simultaneously, the sub-surface $\sqrt{5}$-oxide site in which the oxygen atom resided before the patch formation is deactivated by setting its occupation to {\tt null}.

If the new patch lies next to already existing patches, CO diffusion from neighboring Pd(100) sites onto bridge sites of the new Pd(100) patch is possible, if the nearest-neighbor hollow as well as all up to next-nearest-neighbor bridge sites are empty -- again in accordance with the site-blocking rules for CO diffusion in the Pd($100$) single-lattice 1p-kMC model \cite{hoffmann_co_2014}. If all Pd(100) patch bridge and hollow sites are empty, non-activated CO adsorption is possible onto a bridge site. CO desorption has a desorption barrier of 1.93\,eV and obeys detailed balance with the adsorption process as in the Pd($100$) single-lattice 1p-kMC model. As additional reaction processes we only consider the most likely process labeled E in Fig. \ref{reaction_partners_CO_D} with a barrier of 0.95\,eV, involving a CO molecule adsorbed at site D of the patch and an O atom adsorbed at a directly adjacent upper hollow $\sqrt{5}$-oxide site.

With these prescriptions and new processes the multi-lattice 1p-kMC simulation will allow for the generation of Pd(100) patches, which if generated next to each other will increasingly lead to larger Pd(100)-type domains in the course of a reduction condition simulation. As a Pd(100) unit-cell of the commensurate $\sqrt{5}$-oxide/Pd(100) interface exhibits only 4/5 of the Pd atom density, these patches will not correspond to dense Pd(100) terraces though. Instead, they will contain a regular array of Pd vacancies. In reality, fast Pd self-diffusion will lead to the coalescence into closely packed Pd(100) islands, and connected with it to an entire class of additional elementary processes. However, in the here proposed atomistic pathway, cf. Fig.~\ref{intermediate_structures}, the Pd($100$)-patch formation step shifts the two mobile Pd atoms in such a way that the emerging Pd($100$) vacancy is located furthest from the relevant $\sqrt{5}$-oxide/Pd(100) domain boundary, where the cross-reaction E of Fig. \ref{reaction_partners_CO_D} can take place. As such we anticipate that the coalescence of Pd vacancies through Pd self-diffusion occurs predominantly displaced from the reduction front and then has no effect on the actual reduction kinetics. Within this view, we neglect Pd self-diffusion in the multi-lattice 1p-kMC model, such that the end product of a complete reduction of the $\sqrt{5}$-oxide film corresponds to a Pd(100) surface with a regular array of Pd vacancies.

\begin{table}
\caption{DFT parameters entering rate constants of elementary processes near Pd(100)/PdO$(\sqrt{5}\times \sqrt{5})R27^{\circ}$ domain boundary. All energies are in eV.
}
\begin{tabular}[t]{l|r}
process  & barrier \\
\hline
oxide reconstruction & 0.0 \\
O spillover & 0.66 \\
O reverse spillover & 0.77 \\
O hollow diffusion near nucleus & 1.1 \\
cross-reaction (E) & 0.95 \\ 
\end{tabular}
\end{table}

\subsection{Multi-lattice 1p-kMC simulations of oxide reduction}

\begin{figure}
\includegraphics[width=5cm]{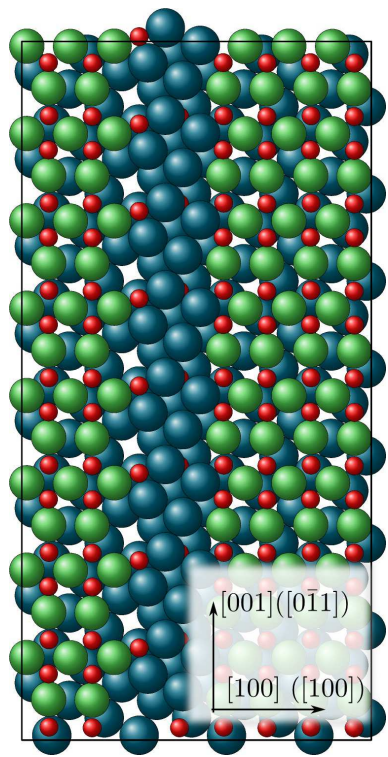}
\caption{\label{initial_oxide_snapshot}Illustration of the initial configuration used in the multi-lattice 1p-kMC simulations. For clarity, only a small $(4 \times 8)$ commensurate $\sqrt{5}$-oxide cell area of the larger simulation cell is shown. One row of $\sqrt{5}$-oxide cells has been reduced to Pd(100) nuclei along the (100) direction of the multi-lattice kMC model. Shown in parenthesis are the corresponding crystallographic axes of the PdO crystal.}
\end{figure}

\begin{figure}
\centering{
\includegraphics[width=8cm]{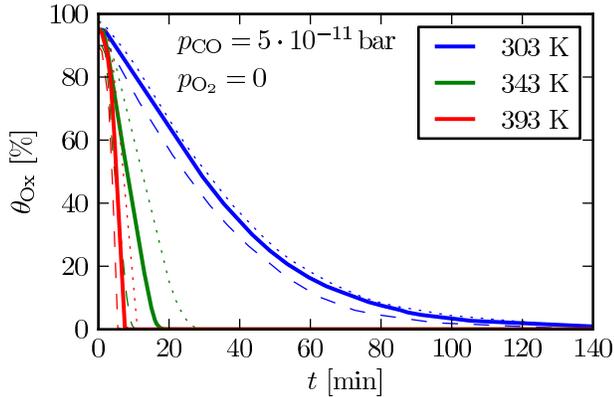}
}
\caption{\label{oxide_averaged_ensemble}Multi-lattice 1p-kMC simulations of the $\sqrt{5}$-oxide reduction at the three experimentally employed temperatures. Shown is the fraction of the surface covered by the surface oxide as a function of time using a $(20 \times 20)$ (solid, bold lines), a $(10 \times 20)$ (dashed, thin lines), and a $(40 \times 20)$ (dotted, thin lines) simulation cell. The slight difference in the initial $\sqrt{5}$-oxide fraction for the different cell sizes results from the fact that in every cell initially one row of oxide is removed.}
\end{figure}

Within the multi-lattice 1p-kMC model we now address the oxide reduction experiments of Fernandes {\em et al.}. The qualitatively new feature introduced by the explicit account of reduced oxide patches in the multi-lattice simulations is the possibility of reduction through cross-reactions at $\sqrt{5}$-oxide/Pd(100) domain boundaries. Within the picture of some initial slow reduction during the long induction time seen in the experiments, we employ kMC simulation cells, in which initially one row of $\sqrt{5}$-oxide along the $[010]$-direction is already removed, cf. Fig.~\ref{initial_oxide_snapshot}. To assess the sensitivity on this point, we conduct the multi-lattice 1p-kMC simulations using $(10 \times 20)$, $(20 \times 20)$ and $(40 \times 20)$ cells, thereby systematically varying the width of the initially  intact $\sqrt{5}$-oxide domain. As demonstrated in Fig. \ref{oxide_averaged_ensemble} we obtain only small variations in the oxide reduction times. These do not affect at all the conclusions put forward below and confirm the understanding that the oxide reduction rates seen in experiment are governed by a fast reduction following some slow activation during the induction period.

\begin{figure}
\centering{
\includegraphics[width=8cm]{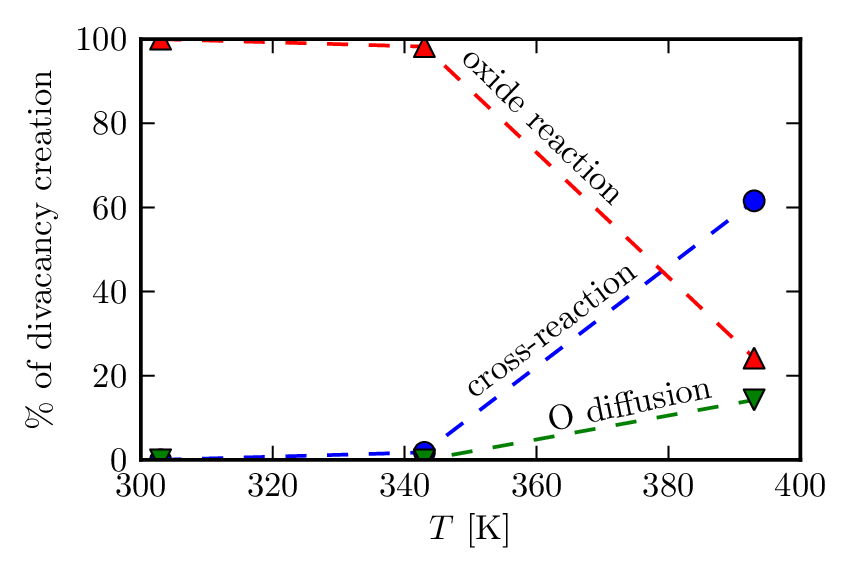}
}
\caption{\label{diva_triggers}
Relative contribution of elementary processes leading to the creation of divacancies and therewith to the ongoing oxide reduction through the formation of Pd(100) patches at the surface. "Oxide reaction" denotes divacancy creation through a CO oxidation reaction within the $\sqrt{5}$-oxide domain, "cross-reaction" denotes divacancy creation through a cross-reaction over the $\sqrt{5}$-oxide/Pd(100) domain boundary, and "O diffusion" denotes divacancy creation through diffusion of O atoms on the $\sqrt{5}$-oxide. The data is obtained using a $(20 \times 20)$ simulation cell.}
\end{figure}

Figure~\ref{oxide_averaged_ensemble} shows the surface oxide coverage as a function of time using the experimental conditions of $5\times 10^{-11}~{\rm bar}$ CO pressure, and temperatures of 303~K, 343~K, and 393~K. In contrast to the single-lattice 1p-kMC simulations discussed in Section~\ref{singlat_sim}, very good agreement with the experimentally observed times and trends is obtained -- without the need to modify any of the 1p rate constants. In order to better understand how the new cross-reaction feature leads to this dramatic improvement, we analyze which elementary process immediately precedes the formation of an appropriate divacancy and therewith triggers the creation of a Pd(100) patch in the simulations. Figure~\ref{diva_triggers} shows these relative contributions and reveals that at lower temperatures reduction is predominately caused by CO oxidation processes within the surface oxide domain, while at the higher temperatures the ongoing reduction is primarily driven by cross-reactions over $\sqrt{5}$-oxide/Pd(100) domain boundaries. The latter is a natural consequence of the enhanced CO binding at the Pd(100) nuclei as compared to the $\sqrt{5}$-oxide. At the higher temperatures CO is then predominantly stabilized on the Pd(100) patches at the surface, concomitantly enhancing the cross-reaction compared to the regular CO oxidation within the surface oxide domain. This change in the mechanism behind the oxide reduction leads to a temperature dependence of the reduction time that is impossible to grasp within a single-lattice model, regardless of how one tries to tweak the underlying rate constants.

\section{Summary and Conclusions}

We presented a general multi-lattice 1p-kMC approach that allows to describe surface morphological transitions within lattice kMC, as long as these transitions occur between structures exhibiting commensurate lattices. This approach allows to evaluate long-time kinetic behavior subject to corresponding transitions and is efficient enough to be based on 1p rate constants. Additionally required compared to traditional single-lattice 1p-kMC is a detailed atomistic pathway for the transition, which then needs to be mapped into the 1p-kMC model in form of new elementary processes.

We illustrated the approach by addressing recent oxide reduction experiments at a Pd(100) model catalyst by Fernandes {\em et al.} \cite{fernandes_reduction_2014}. Single-lattice 1p-kMC simulations monitoring the O coverage as an oxide stability indicator can not account for the temperature dependence of the reduction time observed in the experiments. This holds, even when abandoning the first-principles character of the simulations and optimizing the rate constants of relevant elementary processes to fit the experimental data. In particular at more elevated temperatures, the limitations in stabilizing CO at the surface oxide lead to reduction times that are much too long. This leads us to conclude that the continued oxide reduction subsequent to a long induction period seen in the measurements is not governed by processes at intact oxide patches.

Extensive DFT calculations were correspondingly used to establish an atomistic reduction pathway that generates local Pd(100) patches. An important rational in this development was the realization that several oxide metal atoms do not necessarily have to change their geometric position during the transition from oxide to pristine metal. The choice to only consider pathways where corresponding position changes are not taking place then leads to a drastic reduction in the space of possibly stable intermediates and could also be a useful strategy to conceive complex reconstruction mechanisms in other systems.

Very good agreement with the measurements is obtained with the multi-lattice 1p-kMC simulations on the basis of the proposed oxide reduction pathway. The crucial new feature leading to this improvement is the possibility for CO oxidation reactions across $\sqrt{5}$-oxide/Pd(100) domain boundaries. These cross-reactions constitute a largely varying contribution to the overall oxide reduction within the temperature range covered by the experiments, and therewith generate a temperature-dependence of the reduction rate that cannot be effectively described by a single-lattice model.

In the catalysis context an interface controlled reactivity is often emphasized. As long as phases with commensurate lattices are involved, multi-lattice 1p-kMC offers the exciting perspective to scrutinize this perception within comprehensive microkinetic modeling. The present work already contributes to this by highlighting that even though reaction barriers may be quite similar on the metal, the surface oxide and across the metal/oxide domain boundary, the latter could still be the dominant reaction mechanism due to the steady-state coverages around the metal/oxide interface. For the present model to directly address steady-state CO oxidation activity primarily additional elementary processes are required that describe surface oxidation, in particular dissociative oxygen adsorption at highly O-precovered surfaces e.g. featuring Pd(100) patches surrounded by $\sqrt{5}$-oxide.

\section*{Acknowledgements}
We gratefully acknowledge support from the German Research Council (DFG) and the TUM Faculty Graduate Center Chemistry, as well as generous computing time at the Supercomputing Center of the Max-Planck-Society, Garching.

\appendix

\end{document}